\documentclass[10pt,preprint]{emulateapj}

\usepackage{amssymb}
\usepackage{amsmath}
\usepackage{wasysym}

\shorttitle{Water vapor in a transition disk}
\shortauthors{Salyk et al.}

\begin{document}

\title{Detection of water vapor in the terrestrial planet forming region of a transition disk}

\author{Colette Salyk}
\affil{National Optical Astronomy Observatory, 950 N Cherry Ave, Tucson, AZ 85719, USA\\
Vassar College Physics and Astronomy Department, 124 Raymond Ave, Poughkeepsie, NY 12604, USA}
\author{John H.\ Lacy}
\affil{Department of Astronomy, The University of Texas at Austin, 2515 Speedway, Stop C1400, Austin, TX 78712, USA}
\author{Matthew J.\ Richter}
\affil{Department of Physics, 1 Shields Ave, University of California, Davis, Davis, CA 95616, USA}
\author{Ke Zhang}
\affil{Department of Astronomy, University of Michigan, 311 West Hall, 1085 South University, Ann Arbor, MI 48109, USA}
\author{Geoffrey A.\ Blake}
\affil{California Institute of Technology, MC 150-21, 1200 E California Blvd, Pasadena, CA 91125, USA}
\author{Klaus M.\ Pontoppidan}
\affil{Space Telescope Science Institute, 3700 San Martin Drive, Baltimore, MD 21218, USA}

\begin{abstract}
We report a detection of water vapor in the protoplanetary disk around DoAr 44 with the Texas Echelon Cross Echelle Spectrograph --- a visitor instrument on the Gemini north telescope.  The DoAr 44 disk consists of an optically thick inner ring and outer disk, separated by a dust-cleared 36 AU gap, and has therefore been termed ``pre-transitional''.  To date, this is the only disk with a large inner gap known to harbor detectable quantities of warm ($T\sim450$ K) water vapor.  In this work, we detect and spectrally resolve three mid-infrared pure rotational emission lines of water vapor from this source, and use the shapes of the emission lines to constrain the location of the water vapor.  We find that the emission originates near 0.3 AU --- the inner disk region.  This characteristic region coincides with that inferred for both optically thick and thin thermal infrared dust emission, as well as rovibrational CO emission.  The presence of water in the dust-depleted region implies substantial columns of hydrogen ($>10^{22}\,\mathrm{cm}^{-2}$) as the water vapor would otherwise be destroyed by photodissociation.   Combined with the dust modeling, this column implies a gas/small-dust ratio in the optically thin dusty region of $\gtrsim 1000$.  These results demonstrate that DoAr 44 has maintained similar physical and chemical conditions to classical protoplanetary disks in its terrestrial-planet forming regions, in spite of having formed a large gap.   
\end{abstract}

  
\section{INTRODUCTION}
Detecting protoplanets in their parent disks would allow us to directly study the interactions and relationships between planets and their birth environments. A sub-class of protoplanetary disks termed transition disks remain strong candidates for disks that harbor giant (proto)planets, and thus have generated a significant amount of study. Transition disks were originally identified via a dearth of near-IR excess emission, indicative of depletion of small grains close to the star \citep[e.g.][]{Strom1989, Koerner1993, Calvet2002a}, and a leading explanation for the formation of many of these dust-depleted disks is clearing by tidal interactions with giant planets \citep[e.g.][]{Rice2003, Zhu2011, Dodson-Robinson2011}.  Many transition disks have since been imaged at millimeter wavelengths, confirming the presence of inner dust holes or clearings \citep[e.g.][]{Brown2009, Andrews2011d}.  In addition, there have been several detections of companions in transition disks using sparse aperture masking and/or direct imaging \citep[e.g.][]{Kraus2012, Huelamo2011, Biller2012, Close2014a}.

Transition disks are far from homogeneous, and their inner regions can have complex properties. Near-IR photometry and interferometry show that many transition disks have dust in their inner regions, including some, termed pre-transitional, with optically thick inner rims \citep[e.g.][]{Eisner2006b, Espaillat2010}.  Many transition disks also have azimuthally asymmetric dust and gas structures \citep[][]{Muto2012, vanderMarel2013, Casassus2013a}, suggesting that they are actively being sculpted by planets. But assessing the potential for planet growth, and characterizing planets' birth environments, also requires knowledge of the gas content and chemistry.

\begin{figure*}
\epsscale{1}
\plotone{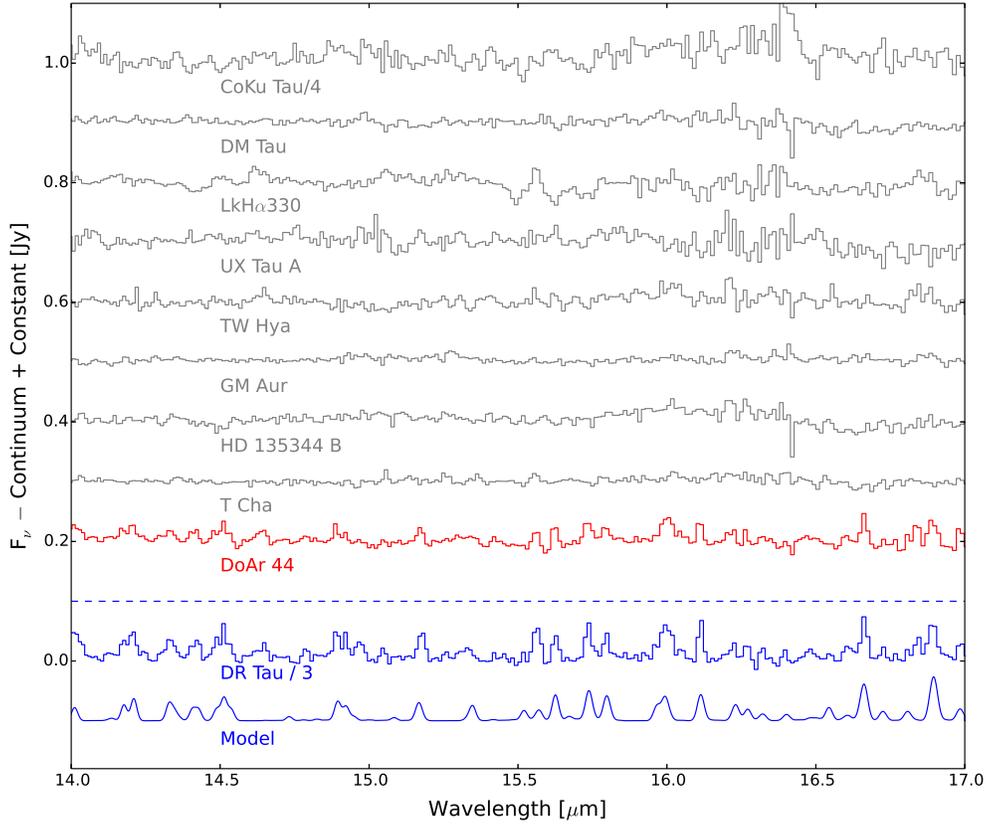}
\caption{Portions of continuum-subtracted {\it Spitzer}-IRS spectra from transition disks, a water-rich disk around DR Tau (divided by 3 for clarity) and an LTE slab model (T=500 K, $N_\mathrm{H_2O}=4\times10^{18}\,\mathrm{cm}^{-2}$ and $R=0.6$ AU).  Transition disk classifications from \citet{Calvet2002a, Furlan2006, Brown2007, Espaillat2010}.}
\label{fig:irs_plot}
\end{figure*}

A powerful probe of conditions in the inner regions of transition disks is water vapor.  Water vapor is easily photodissociated by FUV radiation from young, accreting stars.  Therefore, the presence of water vapor requires shielding of UV radiation and/or sufficiently high gas densities for chemical production \citep[e.g.][]{Adamkovics2014, Du2014}, with possible modifications due to mixing \citep[e.g.][]{Ciesla2006, Albertsson2014}. {\it Spitzer}-IRS spectra of nine transition disks (reduced according to the procedure described in \citealp{Pontoppidan2010b}), a water-rich classical disk, and a water vapor model, are shown in Figure \ref{fig:irs_plot}, demonstrating that nearly all transition disks show no evidence for warm water vapor. Given that $\sim$40\% of young low-mass stars with disks show water vapor emission \citep{Pontoppidan2010b}, it is unlikely that these transition disks are collectively water-poor through chance alone.  Instead, the clearing of the inner regions of transition disks likely reduces dust-induced UV shielding and gas densities which, combined, ``dry out'' the inner disk \citep{Zhang2013}.  In fact, the {\it Spitzer}-IRS spectrum from TW Hya does show water vapor emission at longer wavelengths, but this is emitted by cooler water vapor located at the inner edge of the optically thick outer disk, where gas and dust densities return to higher values \citep{Zhang2013, Kamp2013}.  Therefore, water must already be incorporated into planetesimals, or be delivered, if water-rich planets are to be produced in the inner disk.

An interesting exception to this trend is the disk around the star DoAr 44 (also called ROX 44 and Haro 1-16).  This disk was classified as pre-transitional by \citet{Espaillat2010}, with a spectral energy distribution and near-IR spectrum consistent with the presence of an optically thick ring near 0.25 AU, optically thin dust out to 2 AU, and a cleared gap between 2 and 36 AU.  Its {\it Spitzer}-IRS spectrum shows clear evidence for water vapor (Figure \ref{fig:irs_plot}), with a H$_2$O/CO line ratio similar to classical disks \citep{Salyk2011a}, suggesting a similar C/O ratio in the emitting region.  Given the high upper level energies (up to 6000 K) for the observed water lines, they likely arise from gas close to the young star --- a hypothesis that can be confirmed using spectrally resolved observations.  The presence of water vapor in this disk presents us with an opportunity to better understand the conditions in the inner regions of a disk that has likely already formed giant planets, and what implications this may have for still-forming terrestrial planets.

In this work, we present spectrally resolved observations of water vapor from DoAr 44 and discuss their implications.  In Section \ref{sec:observations}, we discuss the observations and the detection of water vapor.  In Section \ref{sec:results},  we estimate the radial origin of the emission, the water vapor temperature and column density, and the implied hydrogen column density.  In Section \ref{sec:conclusions}, we present our conclusions and discuss their implications.

\section{OBSERVATIONS AND DATA REDUCTION}
\label{sec:observations}
Our target is the protoplanetary disk around DoAr 44 --- a young K3 star in the $\rho$ Ophiuchus cloud \citep{Bouvier1992}.  Spectra were obtained with the Texas Echelon Cross Echelle Spectrograph (TEXES; \citealp{Lacy2002}) --- a visitor instrument on Gemini north --- on 2014 August 15 and 16 UT (Program ID: GN-2014B-Q-40).  The total target integration times were 56 and 39 minutes, respectively, on the two nights.  Observations were performed in the ``hi-medium'' cross-dispersed mode, with a resolution of R$\sim$80,000 at a wavelength of $\sim$12.4$\,\mu$m.  We measure a FWHM of 3.5$\pm0.1\,\mathrm{km}\,\mathrm{s}^{-1}$ (R$\sim$85000) for a narrow sky absorption feature, confirming that we achieve the expected resolution.  The slit size was 0.52$''$ $\times$ 4$''$, enabling nodding along the slit.  

Spectra were reduced utilizing the TEXES pipeline.  Reduction includes flat fielding, subtraction of nod pairs and optimal extraction of spectral traces.  The pipeline derives the wavelength solution from a single atmospheric line in the setting combined with knowledge of the instrument optics, and solutions are found to be accurate to $\sim0.5\,\mathrm{km}\,\mathrm{s}^{-1}$.  To correct for telluric absorption, the bright asteroid Eunomia was observed at a similar airmass, and the spectra from DoAr 44 were divided by this spectrum.  For absolute flux calibration, we scale to the 0.55 Jy 12.4$\,\mu$m continuum flux level measured by \citet{Pontoppidan2010b} in each order.    Uncertainty in absolute flux calibration for {\it Spitzer}-IRS spectra is $\sim$10\% \citep{Furlan2006}; in addition, transition disks often show variability at the 10\%--20\% level in this wavelength region \citep{Espaillat2011}.  Therefore, there is a 10-20\% uncertainty in absolute flux.

\begin{figure*}
\epsscale{1}
\plotone{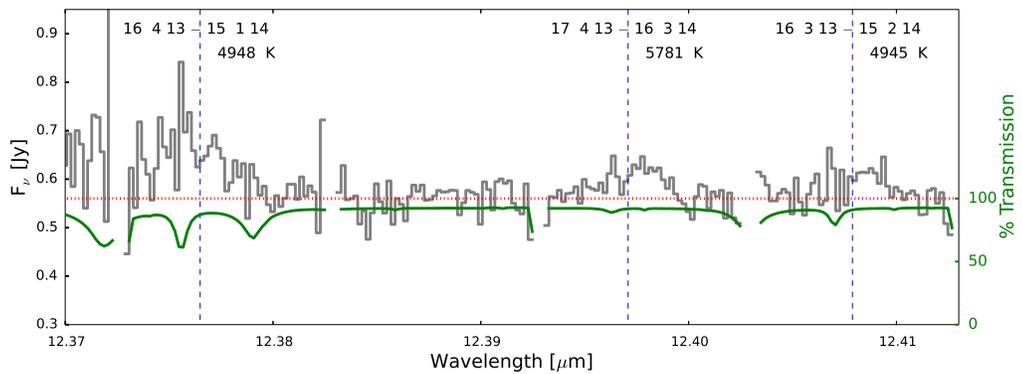}
\caption{Observed spectrum (gray), binned to a resolution of 2 \AA, and atmospheric transmission (green).  Dashed vertical lines mark expected location of water vapor emission.  Labels show upper and lower state quantum numbers (J$_{K_aK_c}$) and upper state energies.  A red dotted line marks the 0.56 Jy continuum.}
\label{fig:spectrum}
\end{figure*}

The full wavelength coverage was 12.36--12.46$\,\mu$m.  Figure \ref{fig:spectrum} shows the portion of the fully reduced spectrum containing three water vapor emission lines expected to emit strongly.  All three water vapor lines are detected, with line/continuum ratios near 0.2.  Details are shown in Table \ref{table:observations}.  Because the emission lines have low line/continuum ratios and are broad, we have binned our spectrum to 2$\times10^{-4}$ $\mu$m ($5\,\mathrm{km}\,\mathrm{s}^{-1}$) --- a factor of $\sim$ 5 larger than the native pixel size of $\sim$4$\times10^{-5}\,\mu \mathrm{m}$ ($0.9\,\mathrm{km}\,\mathrm{s}^{-1}$).   Data are missing in gaps between orders.

Vertical lines mark expected line locations.  Emission line centers are well separated from telluric water, but the wide line profiles from DoAr 44 result in some overlap.  The atmospheric emission here causes a lower signal-to-noise ratio (S/N) on the blue side of the line profiles, which we believe accounts for the observed line asymmetry.  Alternatively, the disk emission lines could be intrinsically asymmetric, with strongly redshifted emission or blueshifted absorption.  However, this source is accreting at a moderate rate ($9\times10^{-9}\, \mathrm{M}_\odot \mathrm{yr}^{-1}$; \citealp{Espaillat2010}) and has no known evidence for an outflow.  In addition, the broad emission lines are similar in shape to observed CO rovibrational emission lines from this source \citep{Salyk2011b}, which show no evidence for asymmetry.   Therefore, we assume here that the lines arise from the protoplanetary disk itself and that any asymmetries are insignificant.  

\begin{deluxetable}{lllll}
\tablehead{\colhead{Line} & \colhead{$\lambda$\tablenotemark{a}} & \colhead{E$_\mathrm{upper}$} & \colhead{Spin} & \colhead{Flux\tablenotemark{b}}\\
\colhead{(J$_{K_aK_c}$)} &$(\mu$m)&(K)&&}\\
\startdata
$16_{4\ 13}\rightarrow15_{1\ 14}$  &12.3765 & 4948 & ortho & $9.25 \pm 0.94$  \\
$17_{4\ 13}\rightarrow16_{3\ 14}$  & 12.3971 & 5781 & ortho & $3.99 \pm 0.94$ \\
$16_{3\ 13}\rightarrow15_{2\ 14}$  &12.4079 & 4945 & para & $3.26 \pm 0.96$ \\
\enddata

\tablecaption{Detected Lines}
\label{table:observations}
\tablenotetext{a}{Theoretical line centers from HITRAN molecular database \citep{Rothman2013}, shifted by the Earth-induced Doppler shift, $\sim28\,\mathrm{km}\,\mathrm{s}^{-1}$, plus the heliocentric velocity of the source, $-8\,\mathrm{km}\,\mathrm{s}^{-1}$, measured from rovibrational CO emission lines \citep{Salyk2011b}.}
\tablenotetext{b}{$10^{-15}$ erg cm$^{-2}$ s$^{-1}$ from Gaussian fit.  Error prescription from \citet{Lenz1992}.  Errors do not include a 10\%--20\% uncertainty in absolute flux.}
\end{deluxetable}

\section{RESULTS}
\label{sec:results}
\subsection{Location of water vapor emission}
\begin{figure*}
\epsscale{1.1}
\plotone{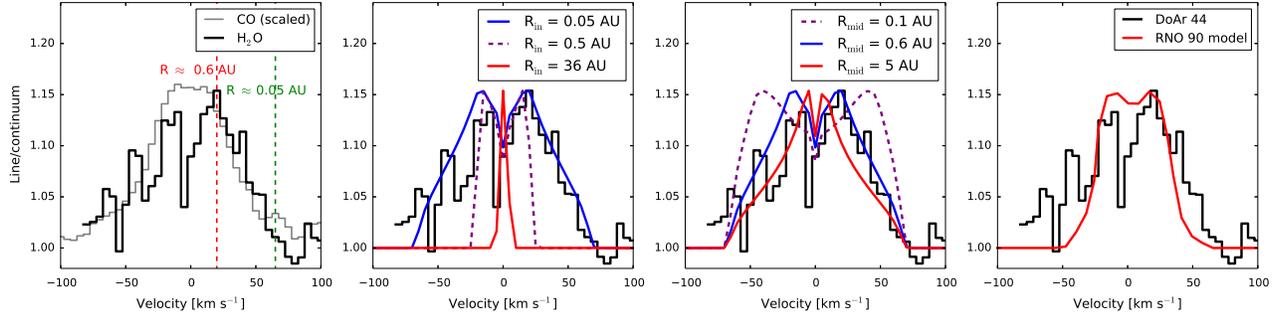}
\caption{{\it First panel:} Weighted average water vapor line profile (black), binned to $5\,\mathrm{km}\,\mathrm{s}^{-1}$, and rovibrational CO line profile (gray; \citealp{Salyk2011b}). Vertical lines mark line peak and line wing (HWZI) velocities, corresponding to outer and inner radii, respectively. {\it Second panel:} Weighted average water vapor line profile (black) and  emission models with different values of $R_\mathrm{in}$.  The models with $R_\mathrm{in}=0.05$ and $0.5$ AU have $R_\mathrm{mid}=0.6$ AU; the model with $R_\mathrm{in}=36$ AU has $R_\mathrm{mid}=40$ AU. {\it Third panel:} Same as second panel, but models have $R_\mathrm{in}=0.05$ and different values of $R_\mathrm{mid}$.   {\it Fourth panel:} Weighted average water vapor line profile (black) and water vapor emission line model (red) from \citet{Pontoppidan2010a} with $R_{in}=0.45$ AU.
}
\label{fig:model_plots}
\end{figure*}

In Figure \ref{fig:model_plots}, we show a weighted (by S/N) average water vapor line profile, as well as an averaged CO rovibrational emission line profile \citep{Salyk2011b}.  While there are differences between the two line profiles, the relatively low S/N of the water emission precludes ascribing this to true differences in the emitting location of the molecules.  Instead, we focus on the primary properties of the emission line --- namely, that it is broad, and possibly double-peaked.  The width of the emission line is clearly inconsistent with arising from the outer disk beyond 36 AU, as the maximum observed velocity in such a case would be only $2.5 \,\mathrm{km}\,\mathrm{s}^{-1}$.  Throughout this work, we assume a stellar mass of 1.4 M$_\odot$ \citep{Andrews2009} and a disk inclination of 25$^\circ$, based on an elliptical fit to the dust emission map in \citet{vanDishoeck2015}.

The peak emission and half width at zero itensity (HWZI) have corresponding velocities of 20 and $65 \,\mathrm{km}\,\mathrm{s}^{-1}$.  Assuming a $10\,\mathrm{km}\,\mathrm{s}^{-1}$ uncertainty, these imply emission arising from an annulus with inner radius $0.05^{+0.07}_{-0.02}$ AU and outer radius $0.6^{+1.7}_{-0.3}$ AU.  Following the procedure outlined in \citet{Salyk2011b}, we also compare the observed lineshape to simplified disk models in which the line luminosity, $L$, is modeled as $L=\int_{R_\mathrm{in}}^{R_\mathrm{out}} L(R) dR$.  We assume $L(R)\propto R^{-1.5}$ from $R_\mathrm{rin}$ to $R_\mathrm{mid}$ and $L(R)\propto R^{-3}$ from $R_\mathrm{mid}$ to $R_\mathrm{out}$, such that the emission drops rapidly beyond $R_\mathrm{mid}$.  We take $R_\mathrm{out}$ as 100 AU.  

Figure \ref{fig:model_plots} demonstrates that a model with $R_\mathrm{in}=0.05$ AU and $R_\mathrm{mid}=0.6$ AU is consistent with the data.  It also shows how the model varies with $R_\mathrm{in}$ and $R_\mathrm{mid}$, and conclusively demonstrates that the observed lineshape is inconsistent with very large values of $R_\mathrm{in}$.  Therefore, the water vapor emission arises in the inner disk.  

Models in which the emission falls more rapidly with radius (i.e., with a steeper L(R) profile) produce the bulk of their flux near $R_\mathrm{in}$, and the observed double-peaked profile could be produced by such a model with slightly larger $R_\mathrm{in}$.   A model of this type, with an inner radius of 0.45 AU, is utilized by \citet{Pontoppidan2010a} to fit water vapor emission from the circumstellar disk around the classical T Tauri star RNO 90, and also provides a reasonably good fit to the observed emission from DoAr 44  (see Figure \ref{fig:model_plots}).  Correcting for RNO 90's higher inclination of 37$^\circ$, the implied inner radius for DoAr 44 would be $\sim$ 0.22 AU.

In summary, we find that the bulk of the line luminosity emerges near 0.3 AU, either from an annulus extending from $\sim$0.05--0.6 AU, or from a region with an inner radius of 0.22 AU.  In the former case, the inner radius is close to the $\sim$0.04 AU sublimation radius (assuming a sublimation temperature of 1500 K and $L=1.3 L_\odot$; \citealp{Andrews2009}).  The two alternative models may be distinguished with higher S/N data by the depth of the line profile's central depression.  

\subsection{Water temperature and column, and implications for local conditions}
\begin{figure*}
\epsscale{1}
\plotone{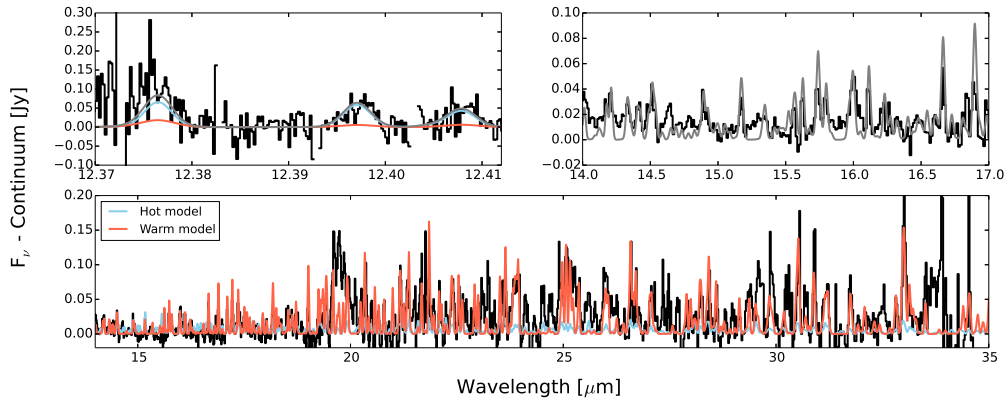}
\caption{{\it Top left:} Continuum-subtracted TEXES spectrum (black), ``hot'' model fit to TEXES and 14-17$\,\mu$m {\it Spitzer} data (blue), ``warm'' model fit to {\it Spitzer}-IRS data from \citet{Salyk2011a} (red) and sum of models (gray).
{\it Top right:} Continuum-subtracted {\it Spitzer}-IRS spectrum (black) and combined hot and warm models (gray).
{\it Bottom:} Continuum-subtracted {\it Spitzer}-IRS spectrum (black), hot model (blue) and warm model (red).
 }
\label{fig:slab_plot}
\end{figure*}

We investigate the properties of the water vapor column by considering it as a single-temperature slab of gas in LTE, with a column density, $N_\mathrm{H_2O}$, area $\pi R^2$ (where we refer to $R$ as the emitting radius), and temperature, $T$ (and assuming a distance of 125 pc and turbulent velocity of $2\,\mathrm{km}\,\mathrm{s}^{-1}$).  The TEXES emission lines alone are not constraining, so we consider them in conjunction with the {\it Spitzer}-IRS data \citep{Salyk2011a} over the 14-17$\,\mu$m range.  The combined fit requires $T\gtrsim500$ and $N_\mathrm{H_2O}\gtrsim10^{18}\,\mathrm{cm}^{-2}$, with the lowest temperatures requiring higher $N_\mathrm{H_2O}$ and vice-versa.  A representative model has $N_\mathrm{H_2O}=10^{19}\,\mathrm{cm}^{-2}$, $T=800\,$K and $R=0.15$ AU (the ``hot'' model in Figure \ref{fig:slab_plot}).

A fit to the full {\it Spitzer}-IRS spectral range found $T=450\,\mathrm{K}$, $N_\mathrm{H_2O}=2\times10^{18}\,\mathrm{cm}^{-2}$, and $R=0.79$  AU \citep{Salyk2011a}.  This ``warm'' model (Figure \ref{fig:slab_plot}) underpredicts the TEXES line fluxes by a factor of 10.  Conversely, the single-temperature model fitting the TEXES data does not produce enough long wavelength flux to match the {\it Spitzer} data beyond $\sim$17$\,\mu$m.  This discrepancy suggests that a range of temperatures is required to explain the observed water vapor emission.

The modeled column density, $N_\mathrm{H_2O}\gtrsim10^{18}\,\mathrm{cm}^{-2}$ and implied $\mathrm{H_2O}/\mathrm{CO}$ ratio of $\gtrsim$1 \citep{Salyk2011a} are consistent with those found for classical protoplanetary disks \citep{Carr2011a, Salyk2011a}.  This column of water implies shielding from photodissociating FUV radiation.  Whether this shielding is provided primarily by dust grains or by self-shielding depends on the dust opacity, with self-shielding dominating when FUV opacities drop below 1\% of canonical ISM values \citep{Bethell2009a}. According to the models of \citet{Espaillat2010}, the water vapor emitting region overlaps both an optically thick dust ring (with radius $<0.4\,$AU) and an optically thin dusty region within 2 AU, whose vertical optical depth is only 0.036 at 10$\,\mu$m.  For 0.1$\,\mu$m pure silicate dust grains with optical constants from \citet{Ossenkopf1992}, the corresponding optical depth at a photodissociating wavelength of 110 nm is 0.56.   The actual optical depth will depend on the precise composition and dust grain size.  However, if grains have grown beyond 0.1$\,\mu$m in size, the UV opacity will be lower.  Therefore, the dust is unlikely to provide sufficient opacity to shield the H$_2$O in this region.

If the dust optical depth is low, models of self-shielding \citep{Bethell2009a} relate observed water column density to vertical hydrogen column density.  For an example disk around AA Tau, with an accretion rate of $\sim3\times10^{-9}\, M_\odot\, \mathrm{yr}^{-1}$ \citep{Gullbring1998}, a water column density of $10^{18}\,\mathrm{cm}^{-2}$ (the minimum column density required to fit the combined TEXES and {\it Spitzer} spectra) corresponds to a vertical hydrogen column of $\sim10^{22}\,\mathrm{cm}^{-2}$, for a range of temperatures ($T\gtrsim600\,$K).  DoAr 44's slightly higher accretion rate of $9\times10^{-9} M_\odot\, \mathrm{yr}^{-1}$ (and correspondingly higher accretion luminosity) implies a factor of a few larger hydrogen column.  In addition, if the water vapor emission lines are sub-thermally excited \citep{Meijerink2009}, required water vapor column densities and hence hydrogen column densities will be higher still.  Therefore, the water vapor detection implies that it arises from a region with a vertical hydrogen column $N_\mathrm{H}>10^{22}\,\mathrm{cm}^{-2}$.  These models assume a canonical water abundance, but true abundances may differ from this value by factors of 5--10, depending on relative rates of gas and icy body radial transport \citep{Ciesla2006}.   If planets form DoAr 44's gap, these may block the transport of icy bodies, resulting in a lower water abundance and higher implied hydrogen column.

If the water vapor emitting region overlaps with the optically thin region producing the 10$\,\mu$m dust emission \citep{Espaillat2010}, we can estimate the gas/(small)-dust ratio in this region.  For 0.1--1$\,\mu$m sized silicate grains, the opacity at 10$\,\mu$m is $\sim2100\,\mathrm{cm}^{2}\,\mathrm{g}^{-1}$ \citep{Ossenkopf1992}, so an optical depth of 0.036 implies a dust mass surface density of $1.7\times10^{-5}\,\mathrm{g}\,\mathrm{cm}^{-2}$.  Dividing the hydrogen mass surface density by the dust mass surface density gives a gas/dust ratio of $\sim$1000.  In this estimate, all of the (optically thin) dust is accounted for, but the hydrogen column derived from the models of \citeauthor{Bethell2009a} is that lying above the water-emitting layer, and so represents a lower limit to the total vertical hydrogen column.  Therefore, the gas/(small-)dust ratio in the optically thin dust region is $\gtrsim$1000.  

\section{CONCLUSIONS AND DISCUSSION}
\label{sec:conclusions}
We detect velocity-resolved water vapor emission lines from the pre-transition disk DoAr 44, using the TEXES spectrograph on Gemini north.  We show that:
\begin{itemize}
\item The water vapor emission arises from the inner disk near 0.3 AU --- well inside the $\sim$36 AU inner rim of the optically thick outer disk, and overlapping the dusty inner disk regions modeled by \citet{Espaillat2010}.
\item The water vapor observed with TEXES requires warmer temperatures than are required to fit the {\it Spitzer}-IRS spectrum.  Therefore, the emission arises from a region with a range of temperatures.
\item The best-fit water column density ($N_\mathrm{H_2O}>10^{18}\,\mathrm{cm}^{-2}$) implies a vertical hydrogen column of $N_\mathrm{H}>10^{22}\,\mathrm{cm}^{-2}$, based on models for water self-shielding \citep{Bethell2009a}.  For the optically thin dusty region \citep{Espaillat2010}, this implies a gas/(small-)dust ratio $\gtrsim1000$.  
\end{itemize}

Many transition disks are active accretors \citep[e.g.][]{Najita2007} and harbor CO gas in their inner dust-depleted regions \citep{Rettig2004, Salyk2007} --- facts that support a planet-clearing origin for these disks \citep{Dodson-Robinson2011}.  Here, we furthermore demonstrate that the hydrogen column in the inner disk of DoAr 44 must be $>10^{22}\,\mathrm{cm}^{-2}$ in order to maintain a significant column of water vapor.  We also derive a super-canonical gas/(small-)dust ratio, which may be the result of grain growth --- a process likely required to explain the low opacity throughout the disk gap \citep{Zhu2012}.  

Conversely, this work implies that the majority of transition disks, which do not have detectable warm water vapor emission, have not been able to maintain high inner disk gas columns.  Note that we would not expect to detect water vapor as strong as is observed from DoAr 44 without also detecting emission with the {\it Spitzer}-IRS (for spectra with similar S/N to those in Figure \ref{fig:irs_plot}).   Therefore, if other transition disks harbor water vapor in their inner regions, the amount of water vapor must be significantly less than that observed from DoAr 44.  

Why has DoAr 44 maintained a gas-rich, water-rich inner region, while other transition disks have not?  Although DoAr 44 has been termed pre-transitional, implying that it may be dissipating its inner disk to evolve into a transition disk, the expected timescale for dissipation of the inner disk is short.  The viscous timescale at 0.3 AU for an actively accreting disk like DoAr 44 (with $\alpha=0.01$) is $\sim$1000 years \citep{Ruden1999} --- less than a hundredth of the disk lifetime.  More likely, the inner disk is being replenished by the outer disk, in agreement with recent observations showing that the {\it outer} regions of transition disk gaps are not gas-free \citep{Casassus2013a, vanderMarel2015}.   If DoAr 44 is not evolving into a transition disk, the differences between DoAr 44 and other transition disks may lie in the architectures of the planetary systems they harbor.  The similarity in stellar and disk properties between DoAr 44 and other transition disks \citep{Andrews2011d} suggests that the late-stage chemistry of the terrestrial planet forming region may depend on a stochastic aspect of planet growth/disk evolution.

Continued study of DoAr 44 will shed more light on the formation scenario for transition disks, and on implications for inner disk chemistry.  Higher S/N mid-IR spectroscopy, ideally combined with spectro-astrometry, would place firmer constraints on the radial origin of the water vapor emission, allowing a uniquely powerful view of the terrestrial planet forming region.  Complementary observations of CO gas in the 10's of AU region have already been obtained with the Atacama Large Millimeter Array (van der Marel et al.\ 2015, in preparation).  By combining these sets of observations, we will better understand the overall architecture of the dust-depleted gap, and what this implies for any embedded planets.

\section{Acknowledgements}
C.~S. acknowledges helpful discussions with Joan Najita.  Observations obtained at the Gemini Observatory, which is operated by the 
Association of Universities for Research in Astronomy, Inc., under a cooperative agreement 
with the NSF on behalf of the Gemini partnership: the National Science Foundation 
(United States), the National Research Council (Canada), CONICYT (Chile), the Australian 
Research Council (Australia), Minist\'{e}rio da Ci\^{e}ncia, Tecnologia e Inova\c{c}\~{a}o 
(Brazil) and Ministerio de Ciencia, Tecnolog\'{i}a e Innovaci\'{o}n Productiva (Argentina).

\end{document}